Brief review

# Current status of real-time imaging of antiproton annihilation in biological targets

Zlatko Dimcovski[1,] and Michael Doser[2]

CERN[2], Geneva, Switzerland, Bioscan.ch[1],

zlatko.dimcovski@cern.ch; michael.doser@cern.ch

Abstract

Following a successful pioneering study of the biological effects of antimatter (the AD-4/ACE experiment at CERN), the use of antiprotons in clinical radiotherapy became a very serious possibility. A major part of any future radiotherapy using antiprotons will be real-time imaging of antiproton annihilation in biological targets. In principle, real-time imaging is possible thanks to the secondary particles, neutral pions and charged pions, produced by the annihilation of antiprotons in the cells they come to a stop in. Neutral pions decay dominantly and almost instantaneously (in about $10^{-16}s$) into a photon pair, and one approach to real-time imaging is based on the detection of these high-energy gammas. An alternative is based on the detection of charged pions, which have lifetimes almost a billion times longer than neutral pions. This paper is a brief review of this field of research, from the first real-time imaging achieved by the AD-4/ACE experiment in 2004 to the present day.

## 1. Introduction

Shortly after CERN's low energy antiproton source, the Antiproton Decelerator (AD), was put into operation, for the first time ever, the biological effects of antimatter were studied in a dedicated experiment (2003 to 2013). Running for ten years, the Antiproton Cell Experiment (or AD-4/ACE experiment) at CERN achieved a significant understanding of the physical and biological phenomena associated with the irradiation of biological targets with antiprotons [1-9], the outcome of which was reviewed in [10].

The AD-4/ACE experiment has clarified many of the unknown aspects that are required to understand the interaction between antiprotons and cells, and has resulted in the very real possibility of antiproton radiotherapy.

Of course, any possible technological realization of antiproton cancer therapy starts with an "antiproton factory" that produces antiprotons and ends in the hospital as efficient antiproton radiotherapy to cure patients. The antiproton factory and the hospital must be "connected", either directly by siting them on the same site, or by transportable medical antiproton traps, to ensure antiproton transfer (and subsequent acceleration to clinically relevant energies) to the hospital.

The technology of low energy antimatter production has been very successfully realized at CERN. While sufficient for now, scaling up to the significantly larger antiproton production rates that an eventual antiproton cancer therapy center would require involves planning a dedicated antimatter factory for medical purposes (producing many more antiprotons than the CERN factory).

In proton therapy hospitals, at the end of the technological line, a good part the equipment to treat patients with antiprotons would be available; the existing equipment used for protons could be adapted in a relatively straightforward manner. But beyond this obvious technological solution, there are alternative ideas that could lead to better, cheaper and more efficient clinical treatment.

Several technological challenges in between the antimatter production facility and the final irradiation facility at a hospital need to be tackled. Assuming that the production facility and the radiotherapy infrastructure are separate, the first and obvious challenge is the development of transportable medical antiproton traps. A second challenge (which is the subject of this review) is the development of real-time imaging to determine the exact location of the cancer cells killed by antiproton radiation. Knowing which cells have already been sufficiently irradiated and thus killed is invaluable in making the radiation less invasive and less damaging to healthy cells.

In principle, real-time imaging is possible because of the secondary particles, neutral pions ($\pi^0$) and charged pions ($\pi^\pm$), produced by the annihilation of antiprotons at the point where they come to rest in cancerous tissue within the irradiated patient.

Neutral pions have an extremely short lifetime of about $10^{-16}$ seconds; their highly dominant decay is in two photons. The detection of these high-energy gammas, produced in the decay of neutral pions, is the first possibility for the development of real-time imaging, as the photons have a very high probability of leaving the patient's body. In section 2, we review the current status of this type of real-time imaging.

In section 3 we review the current status of the second possible type of real-time imaging, based on the detection of charged pions which, because of their much longer lifetime and low interaction probability, can escape the patient's body with only minimal interactions, before decaying a few metres from the point of annihilation.

Section 4 is an outlook, pointing out, among other things, that after the purely fundamental research completed by the AD-4/ACE experiment, there is a need for a new, this time applied, experiment at CERN that can - in addition to addressing a number of open biological questions that need to be resolved for a full understanding of the interaction between antiprotons and cancerous cells - perform two crucial technological tasks: the construction of a transportable medical antiproton trap that can ensure very low-loss transport of a very large number of antiprotons ($10^{10}$ ~ $10^{12}$) and the development of real-time imaging of clinical quality. With the existence of a transportable medical antiproton trap and real-time imaging, the technological knowledge for antiproton cancer therapy will be nearly complete.

## 2. Real-time imaging based on the detection of gamma rays from neutral pions or antiproton-induced nuclear fragmentation.

The antiproton beam available from CERN to AD-4 was extracted directly from the AD [reference]; the beam consisted of several $10^7$ antiprotons extracted at 500 MeV/c and formed a quasi-instantaneous pulse (pulse length of around 500 ns); all $10^7$ antiprotons thus annihilated almost simultaneously and produced a single burst of several $10^7$ secondary $\pi^0$'s and $\pi^\pm$'s. Imaging such a large number of simultaneously appearing particles is extremely challenging given the impossibility of identifying all individual particles; any technique to image the source of the annihilation had to rely on linearly extrapolating remotely (to reduce the local secondary particle flux) determined particle directions. The AD-4 dummy consisted of a tissue-equivalent cylinder of gelatin within which living V79 hamster cells were suspended, surrounded by a water tank equivalent to a patient's dimensions; the cylinder length was chosen such that the 500 MeV/c antiprotons would stop at around half of its length, at a depth of around 10cm, after having penetrated the gelatin. The intentionally spread-out Bragg peak was expected to have a width of the order of several mm's, within which the bulk of the antiprotons should annihilate. A very small fraction of antiprotons annihilated in the entrance channel due to in-flight annihilation with atomic nuclei; the challenge was thus to reconstruct the position and width of the Bragg peak from the produced secondary particles with an error of the order of a mm.

Neutral $\pi^0$'s produced in annihilations of antiprotons in cells decay into two 67.5 MeV photons (boosted by the $\pi^0$ momentum and their emission angle) with very low absorption rates in 10 cm water equivalent (in the order of 10%). Reconstructing their direction at the position of a detector is challenging: typical detectors can only stop such photons, so that directional measurement requires collimation of the (extended) source: a series of parallel collimation gaps (restricting the emission region to the gap size of e.g. 1mm) with flux monitors immediately behind them can map out the emission intensity perpendicular to the slit array with the gap size.

The feasibility of real-time imaging of antiproton annihilation in biological targets ( both V79 hamster cells ( 1-3) and human cells (11) using this technique was first demonstrated in 2004 and 2005 ( 1-3, 11) . This was achieved using IRIS (Interactive Radiotherapy Imaging System), a position sensitive amorphous silicon detector (invented and built by Z.Dimcovski, the co-author of the current paper), which prior to AD-4/ACE was routinely used for high precision medical imaging using 10 – 20 MeV gamma's in hospitals .

IRIS is a patented detector (see bioscan.ch) within a system developed for medical on-line self-triggering imaging applications by BioScan[1] S.A in Geneva, in a cooperation between CERN and HUG (Hopital Universitaire de Genève). It consists of: (a) a converter/scintillator which emits photons when traversed by high energy gammas; (b) a large area flat panel a-Si:H detector matrix which detects photons of visible light with high efficiency; (c) a fast real-time electronic system for readout and digitization of images, protected from radiation damage by its peripheral layout and additional shielding; and (d) appropriate computer tools for control, on-line and off-line analysis, reproduction of images and network transfer.

In order to reconstruct a 3D image of the energy deposition profile using IRIS, a shadow mask was placed between the annihilation vertex and the detector. This mask allows only those gamma's to reach the detector which travel parallel to the mask's channels, therefore producing a true 2D shadow image of the source of the particles. Using several 2D images from different observation angles one can, in principle, reconstruct the full 3D image.

Initial tests of this concept using a simple slit collimator and a spread-out stopping distribution (Figure 1) were successful, but as this attempt was mainly a proof-of-principle within the AD-4/ACE experiment, all but the demonstration of feasibility was left for a possible future experiment focused on the development of technology for antiproton cancer therapy.

---

[1] https://www.bioscan.ch/company.html
https://worldwide.espacenet.com/publicationDetails/biblio?CC=US&NR=6552347&KC=&FT=E&locale=en_EP

Alternatively, detection of prompt gamma radiation released after nuclear fragmentation events initiated by antiproton annihilations should result in detectable prompt x-rays (this time in the 10's of keV ~ few MeV range, the former stemming from transitions in the formed antiprotonic atoms, the later from nuclear processes in the nuclear remnant after antiproton annihilation at its surface); the resulting resolution on the ion range is of the order of 1 mm [12]. As for $\pi^0$'s, a collimation array is needed [13] but given the much lower energies, such a device will be much more effective. On the other hand, the absorption probability of such lower energy gammas in the irradiated body will be much greater than for 67.5 MeV photons, and furthermore, Compton scattering will also lead to unwelcome backgrounds and image smearing. This effect can also be used for imaging detectors, leading to Compton cameras [14] as an alternative with comparable spatial resolution to detectors relying on multi-slit collimation.

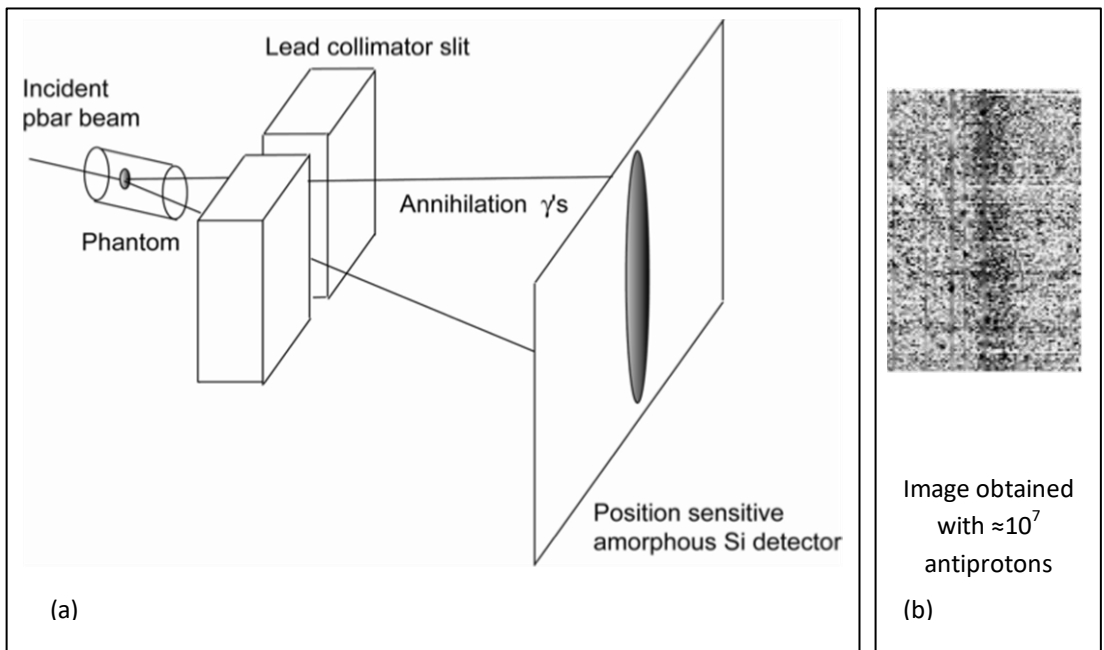

(a) (b) Image obtained with ≈$10^7$ antiprotons

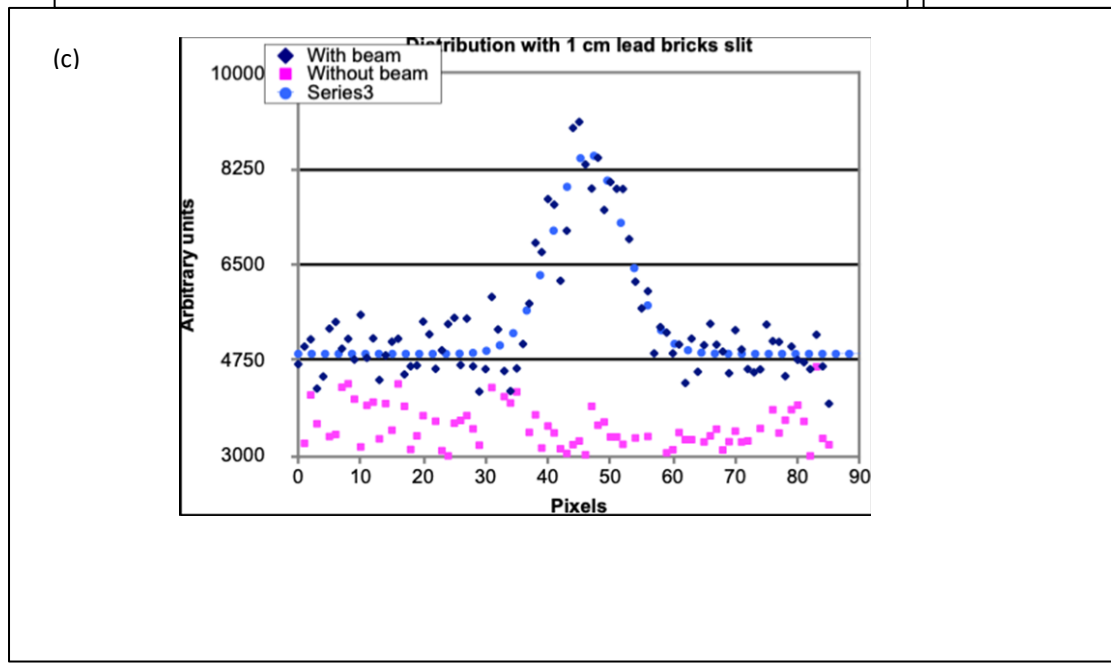

(c)

Figure 1: (a) Schematic set-up for testing the use of high energy gammas for real time imaging. (b) single shot image obtained using IRIS, the position sensitive amorphous silicon detector from BioScan, S.A. with a 1 cm slit between source and detector (c) Counts/pixel for an integration time of 200 ms showing the level of electronic noise and the signal-to-noise ratio obtained with 1 x $10^7$ antiprotons in a single shot of 300 ns duration.

## 3. Real-time imaging based on the detection of charged pions

To reconstruct the origin of a huge number of simultaneously produced $\pi^{\pm}$ particles required a different technique than the traditional approaches of sampling each charged particle's trajectory repeatedly and extrapolating it backwards to the origin, which fail due to combinatorial background. Instead, placing two high resolution two-dimensional detectors at a large distance from the annihilation point. The local track density was sufficiently low there that the assumption that the outer detector's image map and the inner detector's image map could be overlaid, matching hits identified, and linear extrapolation on a hit-pair by hit-pair basis attempted. Multiple scattering of low energy pions in the water dummy could not be corrected for as the rudimentary charged particle tracker had no momentum resolution. First tests of real time imaging were performed using two planes of a prototype chip of the ALICE silicon detector [7, 15].

The first, but unsuccessful, attempt to demonstrate the feasibility of real-time imaging based on the detection of charged pions in spite of the daunting conditions was made by the AD-4/ACE collaboration in the summer of 2004 [1] and again in the summer of 2005 [2, 11]. In fact, it was carried out at the same time as the successful demonstration of the feasibility of real-time imaging based on the detection of gamma rays from neutral pions, as described in the previous section.

Finally, after seven years of persistence the AD-4/ACE collaboration demonstrated the feasibility of real-time imaging using charged pions [7, 15], an experimental success which was strongly supported by Monte Carlo simulations [16, 17]. One limiting fact - the need to position the detector at a sufficient distance from the annihilation stopping point to not saturate the detector upon pulsed antiproton irradiations - could be mitigated by using higher resolution (smaller pixel size) detectors. The pixel size of the detectors used by AD-4 was 50 μm × 425 μm; very recently, pixel detectors with 1.12 μm × 1.12 μm pixels have been exposed to antiprotons, and minimum ionizing charged particles were successfully identified. While maintaining a detector occupancy of less than 8% in the case of the AD-4 (for $3 \times 10^7$ antiprotons in a 500 ns pulse every 90 s) required placing it at 1.4 m from the water phantom, the much higher resolution detector could be placed significantly closer (30 cm from a water phantom). This would result in an occupancy of less than 4% if the same detector geometry as in [15] is used or well below 1 ‰ in the case of a traditional multi-plane tracking geometry, and would correspond to the prerequisites of the simulated vertex distribution along the beam axis of [15]). Such a topology would then result in a multiple-scattering limited (multiple scattering of the pions within the water phantom, magnified by the distance between phantom and detector) stopping distribution reconstruction with a FWHM of approximately 2 cm. This in turn would lead [15] to being able to reconstruct the range of the antiprotons with an uncertainty in the order of a few millimeters already after the very first fraction of an irradiation (relying on $10^8$ antiprotons).

## 4. Outlook

As we have seen, the feasibility of real-time imaging using charged pions even in the case of simultaneous annihilation of a very large number of antiprotons is supported both by Monte Carlo simulations [16, 17] and, more importantly, by preliminary tests [7, 15] of the AD-4/ACE collaboration at CERN. Naturally, slowly continuously extracted antiprotons, or extraction of long sequences of micro-bunches, would greatly reduce the challenge of using charged pions to reconstruct in real time the antiproton stopping distribution and would furthermore allow dynamically adapting the antiproton energies to distribute the deposited dose over a larger region, with real-time imaging feedback ensuring that the the irradiation plan would be fulfilled.

The current state of real-time imaging based on neutral pions is much better matched to simultaneous delivery of very large numbers of antiprotons and is not affected by multiple scattering of the pions within the irradiated body; furthermore, it has been successfully tested by IRIS, a clinical-quality detector used in a number of hospitals. The additional support provided by the Monte Carlo simulations [16], a few years after the experiments [1-3, 11], will help further improving the imaging capability of this mode. In contrast, imaging with charged pions is technically more straightforward, but will suffer from resolution smearing due to multiple-scattering of charged pions between the antiproton stopping point and the detector. Finally, detection of prompt gamma radiation released after nuclear fragmentation events initiated by antiproton annihilations should allow determining ion ranges in real-time with an accuracy of about 1 mm [12].

While the main focus of AD-4/ACE was on an in-depth understanding of the deposited dose, it did also establish the feasibility of real-time anti-imaging; the development of clinical-quality real-time anti-imaging was however far beyond its scope.

The challenging development of clinical quality real-time anti-imaging is only possible within a new project at CERN, which would also allow addressing a number of open questions regarding the suitability of antiprotons specifically for different types of cancerous tissue, as well as in-vivo. Hence, after the success of the AD-4/ACE experiment, which was strongly focused on fundamental science, we need a completely different kind of experiment at CERN, fully focused on future medical applications, of which real-time anti-imaging is one of the most important. Such a new experiment should stand at the beginning of a complex engineering and medical task: the development of a technology line from the antimatter factory to the hospital for antiproton radiotherapy. Two major required technological developments, real-time clinical quality imaging and a transportable medical antiproton trap, can be achieved within such a project at CERN.

Concerning transportable antiproton traps, the current state of the art is two-fold: the PUMA experiment has developed a not yet commissioned transportable device with the goal of combining antiprotons with radio-isotopes (which are not available at the AD); a second transportable antiproton trap has been built by the BASE-STEP collaboration [18], currently demonstrated only with protons, but a necessary tool for the fundamental scientific research of the Baryon Antibaryon Symmetry Experiment (BASE) at CERN [18] which requires transfer of antiprotons to a facility where they can be studied with greater precision than is possible in the AD environment. BASE's achievements in building a transportable antiproton trap can be an excellent starting point for building a corresponding medical antiproton trap.

With the existence of a transportable medical antiproton trap and real-time imaging, several of the most important technological challenges for antiproton cancer therapy will have been tackled.